\documentclass[amsmath,11pt]{article}
\usepackage{amsfonts,amsmath,amssymb,graphics,epsfig}
\usepackage{enumerate}\usepackage{theorem}
\usepackage{multicol}

\date{\empty}

\begin{document}

\title{Law of elasticity and fracture limit of magnetic forcelines under their gravitational deformation}

\author{Panagiotis Mavrogiannis\\ {\small Section of Astrophysics, Astronomy and Mechanics, Department of Physics}\\ {\small Aristotle University of Thessaloniki, Thessaloniki 54124, Greece}}

\maketitle

\begin{abstract}
    Magnetic fields are a very special form of elastic medium. Within astrophysical environments (magnetised stars and protogalaxies) they counteract shear and rotational distortions as well as gravitational collapse. Their vector nature allows for their extraordinary coupling with spacetime curvature in the framework of general relativity. This particular coupling points out the way to study magnetic elasticity under gravitational deformation. In this context, we reveal their law of elasticity, calculate their fracture limit and subsequently argue that they ultimately lose the battle against gravitational contraction of magnetised matter. Two illustrative applications, in a neutron star and a white dwarf, accompany the results.
\end{abstract}

\section{Introduction}

It is well known (mainly from astrophysical studies of magnetised fluids, e.g. see~\cite{Pa} and~\cite{M}, but from relativistic as well~\cite{GT}) that magnetic forcelines behave like an elastic medium under their kinematic (shear or rotational) deformation. Namely, in analogy with a spring under pressure they develop tension stresses resisting their deflection. However, it is less known how to achieve a theoretical description of (elastic) magnetic distortion due to gravity~\cite{T1}-\cite{MT} (namely spacetime curvature within general relativity). In particular, the aforementioned studies have shown that the elastic behaviour in question is expressed through a magneto-curvature tension stress coming from the Ricci identities. In fact, due to their vector nature, magnetic fields present a double coupling with spacetime curvature, not only via Einstein's equations but via the Ricci identities as well. Thus, from a relativistic point of view it has been found out that interestingly magnetic forcelines do not self-gravitate~\cite{GT}, \cite{TMav}, \cite{Mel}. Moreover, they counteract gravitational implosion of a highly conducting fluid and potentially hold it up~\cite{T2}-\cite{Ch}. In reference to this problem and given the elastic behaviour of magnetic fields, one can raise the question concerning the existence of a possible elastic and a fracture magnetic limit. Moreover, if such a fracture limit exists, could one provide an estimation of it for a given magnetised (collapsing) star? Another crucial question associated with the previous ones, is whether magnetic forcelines manage to disrupt gravitational collapse before reaching their fracture limit~\cite{MT}.

Addressing the above questions through an insightful introduction to the gravito-magnetic law of elasticity, basically forms the object of the present piece of work, motivated by~\cite{MT}. In detail, we begin with a brief presentation and mathematical description of the kinematically induced magnetic tension stresses. Then, we focus our attention on the magneto-curvature tension stress and reveal the law of magnetic elasticity under gravitational distortions. Subsequently, we  move on to our principal task which consists of a theoretical calculation of the magnetic fracture limit during the gravitational collapse of magnetised matter. The aforementioned limit, illustrated by two examples, of a neutron star and a white dwarf, is ultimately used to argue that magnetic fields are not able to impede gravitational contraction before being broken. As far as we know, the results (intuitive presentation of the gravito-magnetic law of elasticity, calculation of the magnetic fracture limit and its applications) are new, appearing here for the first time in the literature.

\section{Kinematically induced magnetic tension stresses}

To begin with, let us consider the decomposition of the magnetic 3-D gradient $\rm D_{b}B_{a}$ into its symmetric (trace-free), antisymmetric and trace part. In other words,
\begin{equation}
    {\rm D}_{b}B_{a}={\rm D}_{\langle b}B_{a\rangle}+{\rm D}_{[b}B_{a]}+\frac{1}{3}({\rm D}^{c}B_{c})h_{ab}\,,
    \label{magn-gradient}
\end{equation}
which reveals the individual tension\footnote{Actually the magnetic tension force vector refers to the directional derivative along the field itself. See the following analysis.} components triggered by, and resisting to shape (i.e. $\sigma^{(B)}_{ab}={\rm D}_{\langle b}B_{a\rangle}$), rotational (i.e. $\omega^{(B)}_{ab}={\rm D}_{[b}B_{a]}$) and volume distortions (i.e. $\Theta^{(B)}={\rm D}^{a}B_{a}$) of the magnetic forcelines respectively. Besides, at the magnetohydrodynamic limit (MHD) the tension component opposing to volume expansion/contraction (last term) vanishes (i.e. ${\rm D}^{c}B_{c}=0$ from Gauss's law). In the above ${\rm D}_{a}=h_{a}{}^{b}\nabla_{b}$ is the projected (3-D) covariant derivative operator and $h_{ab}=g_{ab}+u_{a}u_{b}$ (with $g_{ab}$ being the spacetime metric and $u^{a}$ being a timelike $4$-velocity vector) an operator projecting upon the observer's (3-D) rest-space.  The covariant kinematics of the magnetic tension stresses are monitored by the Ricci identities for the magnetic field
\begin{equation}
    2\nabla_{[a}\nabla_{b]}B_{c}=R_{abcd}B^{d}\,,
    \label{4-Ricci}
\end{equation}
where $R_{abcd}$ is the Riemann spacetime tensor. In particular, the timelike part of the above leads to propagation equations for the magnetic shear $\sigma^{(B)}_{ab}$ and vorticity $\omega^{(B)}_{a}=\epsilon_{abc}\omega^{bc}$. On the other hand, its spacelike part leads to divergence conditions (constraints) for the aforementioned quantities. The equations in question, appearing here for the first time-as far as we know-, could prove useful when studying the kinematics of magnetised fluids in various contexts. However, as we do not make any use of those in the present manuscript, we have chosen to place them in the brief appendix~\ref{appA}.  

\section{Gravitationally induced magnetic tension stresses}

In analogy with their deflection due to kinematic effects associated with the fluid's motion, magnetic forcelines counteract their gravitational distortion. Where does the corresponding magneto-curvature tension stress come from? The answer lies in the direct coupling of magnetic fields (as vectors) with spatial curvature via the (3-D projected) Ricci identities,
\begin{equation}
     2{\rm D}_{[a}{\rm D}_{b]}B_{c}=-2\omega_{ab}\dot{B}_{\langle c\rangle}+\mathcal{R}_{dcba}B^{d}\,,
     \label{3-Ricci}
\end{equation}
where $\mathcal{R}_{abcd}$ represents the 3-D counterpart of the Riemann tensor. Note that the aforementioned coupling manifests itself at the second differentiation order.

\subsection{Describing the magneto-curvature tension stress}\label{ssec:magneto-curvature tension}

Let us consider the 3-gradient of the magnetic tension force vector $\tau_{a}=B^{b}{\rm D}_{b}B_{a}$ (the non-zero tension force implies that the magnetic fieldlines are not spacelike geodesics). Employing the 3-D Ricci identities~\eqref{3-Ricci} we arrive at
\begin{equation}
    {\rm D}_{c}\tau_{a}={\rm D}_{c}B^{b}{\rm D}_{b}B_{a}+B^{b}{\rm D}_{b}{\rm D_{c}}B_{a}+2\omega_{bc}B^{b}\dot{B}_{\langle a\rangle}+\mathcal{R}_{dabc}B^{b}B^{d}\,.
\end{equation}
The first three terms involve kinematic effects through eq~\eqref{magn-gradient} whilst the last one can be envisaged as the magneto-curvature tension stress (or the gradient of the magneto-curvature tension component). If $n^{a}$ is the magnetic field direction (i.e. $B^{a}=\mathcal{B}n^{a}$), the term in question can alternatively be written as
\begin{equation}
    s_{ac}=\mathcal{R}_{dabc}B^{b}B^{d}=\mathcal{B}^{2}\mathcal{R}_{dabc}n^{b}n^{d}=-\mathcal{B}^{2}u_{ac}\,,
    \label{Magn-curv-tensor}
\end{equation}
where $u_{ac}\equiv-\mathcal{R}_{dabc}n^{b}n^{d}$ can be envisaged as a kind of strain tensor\footnote{Typically, within conventional elastic mechanics the strain tensor is defined to be a dimensionless symmetric quantity~\cite{LL}. However, here we allow for a non-vanishing anti-symmetric part taking into account any torsional deformation. Furthermore, our strain tensor has inverse square length dimensions in geometrised units.}, describing spatial distortions of the magnetic forcelines (for a commentary on the law of magnetic elasticity under volume gravitational distortions refer to section~\ref{sec:Magnetic-elasticity-law}). Our definition for the strain tensor is metric independent and thus essentially differs from its counterpart~(4.9) in~\cite{CQ}. As for the stress tensor $s_{ab}$, it includes those forces which act against (see the following discussion on the problem of gravitational collapse) spatial curvature and tend to restore the forcelines to their initial state. Overall, the meaning of~\eqref{Magn-curv-tensor} is the following. Due to spatial curvature the magnetic fieldlines are bent and twisted. In analogy with an elastic rod under pressure they react via the restoring stress $s_{ab}$ which increase in proportion to the amount of deformation $u_{ab}$ (Hooke's law of elasticity) and the magnetic density. In fact, when appearing in the kinematic equations for a magnetised fluid, it turns out that the magneto-curvature tension stress depends on the ratio of the magnetic density over the total system's density (i.e. matter and magnetic fields-see eq.~\eqref{el-law2} in the following).

Let us recall that any kind of deformation can be reduced into a sum of a pure shear ($u_{\langle ab\rangle}=\mathcal{R}_{d\langle ab\rangle c}n^{c}n^{d}$), a torsional one (or twisting)\footnote{For the sake of accuracy, vorticity or rotational deformations are included in the shear/shape type of distortions as well.} ($u_{[ab]}=\mathcal{R}_{d[ab]c}n^{c}n^{d}$) and a hydrostatic compression ($(u^{c}{}_{c}/3)h_{ab}=(1/3)\mathcal{R}_{cd}n^{c}n^{d}h_{ab}$). Hence, on splitting the magneto-curvature tension stress into its symmetric trace-free ($s_{\langle ac\rangle}$), antisymmetric ($s_{[ac]}$) and trace part ($s=s^{c}{}_{c}$), we receive its associated component counteracting shape, rotational and volume changes respectively, due to gravity (see appendix~\ref{appA}). Of the aforementioned components we focus here on the last one, namely
\begin{equation}
    s=s^{c}{}_{c}=\mathcal{B}^{2}\mathcal{R}_{bd}n^{b}n^{d}=\mathcal{B}^{2}\left(\frac{2}{3}\rho+\mathcal{E}+\Pi\right)\,,
    \label{elast-law1}
\end{equation}
 where $\Pi\equiv \pi_{ab}n^{a}n^{b}$ and $\mathcal{E}\equiv E_{ab}n^{a}n^{b}$ are the anisotropic stress and the tidal (or electric Weyl) tensor, twice projected along the magnetic direction, respectively. In deriving the above we have employed eqs~\eqref{eqn:app-Σ} and~\eqref{eqn:Sigma-vector} from the appendix. Also, assuming an ideal fluid model, the anisotropic stress terms in the above vanish. Then, of particular interest is that the deformation due to gravitational compression/expansion in~\eqref{elast-law1} is determined by the density  of matter and the tidal tensor projected along the magnetic fieldlines. Note that there is no magnetic input, coming from anisotropic stresses (recall that $\pi^{(\text{magn})}_{ab}=-B_{\langle a}B_{b\rangle}$), in~\eqref{twisting} and~\eqref{elast-law1} whilst there is in~\eqref{s-STF}. In fact, $\Pi^{(\text{magn})}_{a}$=0 (for $n^{a}\parallel B^{a}$) and the magnetic tension exactly cancels out the gravitational magnetic contribution in~\eqref{elast-law1} (e.g. refer to~\cite{TMav}).

\subsection{Magneto-curvature tension and gravitational collapse}\label{ssec:Magneto-curvature-tension}

Now having in hand the expression for the magneto-curvature tension stress, how can we reveal its competitive behaviour towards the corresponding cause of magnetic deformation? An illustrative description can be achieved by making use of the shear, vorticity and volume scalar propagation equations (e.g. see~\cite{TCM}). In each case the associated tension term turns out to have opposite sign to its triggering source. Here being especially interested in the problem of magnetised collapse, we focus on the volume scalar propagation equation, known as Raychaudhuri's equation, in combination with Euler's equation for a magnetised fluid. More specifically, for an ideal fluid the latter reads
\begin{equation}
    (\rho+P+B^{2})\dot{u}_{a}={\rm D}_{a}P-\frac{1}{2}{\rm D}_{a}B^{2}+B^{b}{\rm D}_{b}B_{a}+\dot{u}^{b}B_{b}B_{a}\,,
\end{equation}
where the second and third term in the right hand side split the magnetic Lorentz force into its pressure and tension component respectively. Taking, for simplicity, the divergence of the above under the assumption of partial homogeneity\footnote{In fact, the assumption of partial homogeneity (a standard practice in singularity theorems) is not crucial regarding the core of our reasoning. In the following section we examine the elasticity of the magnetic forcelines under their volume gravitational bending. The cause of the magnetic bending (i.e. total gravitational energy) and its consequence (i.e. magnetic deformation) are explicitly independent of the partial homogeneity assumption.} (i.e. ${\rm D}_{a}\rho\simeq 0\simeq {\rm D}_{a}P$ \text{and} ${\rm D}_{a}\mathcal{B}^{2}\simeq 0$ but ${\rm D}_{a}B_{b}\neq 0$), and taking into account eq~\eqref{3-Ricci}, it turns out that
\begin{equation}
    {\rm D}^{a}\dot{u}_{a}=c^{2}_{\mathcal{A}}\mathcal{R}_{ab}n^{a}n^{b}+2(\sigma^{2}_{B}-\omega^{2}_{B})\,.
    \label{4-accel-div}
\end{equation}
Note that $s^{*}\equiv c^{2}_{\mathcal{A}}(\mathcal{R}_{ab}n^{a}n^{b})$ (with $c^{2}_{\mathcal{A}}\equiv \mathcal{B}^{2}/(\rho+P+\mathcal{B}^{2})$ being the Alfv\'{e}n speed) actually comes from~\eqref{Magn-curv-tensor}, by contracting the indices $a$ and $c$, and dividing by $\rho+P+\mathcal{B}^{2}$ (it would have had the same form if we had assumed inhomogeneous contraction). In other words, it represents the magneto-curvature tension component opposing to volume (curvature induced) distortions of the magnetic forcelines (or of the magnetised fluid). On the other hand, $\sigma^{2}_{B}\equiv{\rm D}_{\langle b}B_{a\rangle}{\rm D}^{\langle b}B^{a\rangle}/2(\rho+P+B^{2})$ and $\omega^{2}_{B}\equiv{\rm D}_{[b}B_{a]}{\rm D}^{[b}B^{a]}/2(\rho+P+B^{2})$ refer to the norms of magnetic tensions counteracting shape (shear) and rotational distortions respectively.\\
Finally, the substitution of ~\eqref{4-accel-div} into the well known Raychaudhuri formula, monitoring the magnetised fluid's volume expansion/contraction, brings the latter into the intuitive form
\begin{equation}
     \dot\Theta+\frac{1}{3}\Theta^{2}=-(R_{ab}u^{a}u^{b}-c^{2}_{\mathcal{A}}\mathcal{R}_{ab}n^{a}n^{b})-2(\sigma^{2}-\sigma^{2}_{B})+2(\omega^{2}-\omega^{2}_{B})+\dot{u}^{a}\dot{u}_{a}\,.
     \label{Raych-mod}
\end{equation}
Each parenthesis in the above includes two opposite sign terms (negative sign terms in the right hand side favour volume contraction whilst positive ones favour expansion~\cite{W}), the cause of magnetic deformation and the associated tension (presented above) opposing to it. Interestingly, we observe that closed spatial distortions ($\mathcal{R}_{ab}n^{a}n^{b}>0$, this is actually the case of stellar gravitational implosion) along the magnetic direction $n^{a}$ give rise to stresses $c^{2}_{\mathcal{A}}\mathcal{R}_{ab}n^{a}n^{b}$ opposing to contraction whilst open spatial distortions ($\mathcal{R}_{ab}n^{a}n^{b}<0$) generate magneto-curvature tension stresses reinforcing contraction. For the rest of this work we focus our attention on the first couple of terms (gravitational deformation of the magnetic field and its elastic reaction) in the right-hand side of eq.~\eqref{Raych-mod}.

\section{The law of magnetic elasticity under (volume) gravitational distortions}
\label{sec:Magnetic-elasticity-law}

With the present section we move on to the essential part of our work. In detail, we discuss the law of magnetic elasticity, describe an enlightening analogy to magnetised collapse, calculate the magnetic fracture limit and apply it to the problem of gravitational collapse of compact stellar objects. The present section should be studied along with appendix~\ref{appB-after}.\\ \\
In reference to equation~\eqref{Raych-mod} we particularly observe that the magneto-curvature tension stress
\begin{equation}
    s^{*}=-c^{2}_{\mathcal{A}}u\,,
    \label{el-law2}
\end{equation}
counteracts the magnetised fluid's gravity, $R_{ab}u^{a}u^{b}=(1/2)(\rho+3P+\mathcal{B}^{2})>0$, namely the cause of magnetic (volume) distortion in the form of $u\equiv-\mathcal{R}_{ab}n^{a}n^{b}$ (so that $u<0$ is associated with closed spatial sections and compression). We plausibly require that $\mathcal{R}_{ab}n^{a}n^{b}>0$ ($u<0$) at all times during gravitational contraction. In complete analogy with~\eqref{Raych-mod}, the symmetric-trace-free and the antisymmetric counterparts of~\eqref{el-law2} can be obtained via the shear and the vorticity propagation formulae respectively, along with Euler's equation of motion. However, here we examine the magnetic elasticity against gravitationally induced (volume) distortions.

\subsection{Insight into the law of magnetic elasticity}

The meaning of~\eqref{el-law2}\footnote{The expression in question has appeared several times in past works (e.g. see~\cite{T2}-\cite{MT}) but it was not recognised or envisaged as an expression of Hooke's law of elasticity and therefore was not given its full interpretation presented here.} is that the tension stress $s^{*}$, tending to restore the magnetic field into its initial (undeformed) state, is proportional to the distortion of the magnetic forcelines (refer to the appendices~\ref{appB-after} and~\ref{appB})\footnote{Written here for an ideal (magnetised) fluid.}
\begin{equation}
u\equiv-\mathcal{R}_{ab}n^{a}n^{b}=-(2/3)\rho-\mathcal{E}\,.
\end{equation}
The minus sign in the right-hand side of~\eqref{el-law2} implies that $s^{*}$ acts against the increasing magnetic distortion $u$. Note that in deriving the above we have taken into account eq. (1.3.41) from~\cite{TCM}, as well as relations~\eqref{eqn:app-Σ}, \eqref{eqn:Sigma-vector} from the appendix~\ref{appB-after}. It is straightforward to see that condition $\mathcal{E}=-(2/3)\rho$ corresponds to the natural (undeformed) `volume' state of the magnetic field, where $s^{*}=0$. The proportionality factor $0<c^{2}_{\mathcal{A}}<1$ (note the difference to~\eqref{Magn-curv-tensor}) is always positive and its definition implies that the greater the magnetic density contribution to the total fluid's density, the more rigid the magnetic fieldlines are (or the more they resist to their deformation). In other words, eq~\eqref{el-law2} is a relativistic expression of Hooke's law of elasticity for a gravitationally distorted magnetic field, frozen into a highly conducting fluid. Nevertheless, in contrast to an elastic spring, the proportionality  factor $c^{2}_{\mathcal{A}}$ is not a constant but a variable quantity (a function of the ratio $\rho/\mathcal{B}^{2}$). Moreover, although Hooke's law is an approximate relation valid for sufficiently small deformations, eq~\eqref{el-law2} seems to be valid for any deformation, given that the Ricci identities~\eqref{3-Ricci} hold. Therefore, from our point of view, magnetic fields appear to keep their elastic behaviour as well as to satisfy Hooke's law of elasticity no matter how big their deformation is.

Even if magnetic forcelines do not present an elastic limit\footnote{The elastic limit refers to that value of distortion beyond which the elastic medium is unable to return to its initial state. Mathematically speaking, on setting the external forces equal to zero, the deformation becomes zero as well. Of course we do not know any such example of material in nature.} under their gravitational bending, one expects that they can support a finite amount of distortion. Thus, we expect that there must be at least a fracture limit of the magnetic fieldlines, predicted by the exact formula~\eqref{el-law2}\footnote{In contrast, Hooke's law for elasticity, being a linear approximation-valid for small values of deformation-, does not and could not contain such information.}. The significance of such a limit becomes clear on considering for instance the astrophysical/cosmological phenomenon of magnetised gravitational collapse. In particular, magnetic fields are known not to self-gravitate as well as to have the potential to impede gravitational implosion from reaching a spacetime singularity. Before proceeding to a definition and theoretical calculation of the magnetic fracture limit under gravitational distortions, we present our approach to magnetised contraction through an analogy.

\subsection{Our approach to magnetised gravitational collapse--An illuminating analogy}\label{ssec:Approach-magn-grav-collapse}

We study gravitational collapse of a magnetised fluid. In other words, we study the contraction of a medium which behaves elastically along a specific direction $n^{a}\parallel B^{a}$. The above statement points out the crucial difference between `ordinary' and magnetised gravitational collapse, which we illustrate via the following analogue.\footnote{The present subsection should be studied along with appendix~\ref{appB-after}}\\
Consider an elastic waterballoon filled with water\footnote{We assume that there is no water or air flow in or out of the balloon.}. In this case water stands for the fluid whilst the elastic medium (balloon containing the water) stands for the magnetic field. Both the balloon and the magnetic field, force, in a sense, the water/fluid to behave elastically. Now imagine that we moderately tight the waterballoon with our hand (the tightening stands for gravitational contraction). We will observe that the waterballoon increases in length along a specific direction while in parallel its $2$-D sections orthogonal to the latter, contract. Overall, the volume of the waterballoon does not change (it contains always the same water volume); the individual dimensions, $1+2$, are those which actually change. Ultimately, if the tightening becomes too big (the external pressure much greater than the elastic restoration force), the balloon breaks releasing the (no longer elastic) water. We expect that the collapsing magnetised fluid should present an analogue behaviour. In fact, we approach the above described elasticity along the magnetic direction $n^{a}$ by adopting the assumption $u'_{a}=n^{b}{\rm D}_{b}u_{a}=0$. (i.e. homogeneous velocity along the magnetic fieldlines-streamlines in this case). In particular, $u'_{a}=0$ (i.e. homogeneous velocity along the magnetic fieldlines) implies that \textit{the magnetic forcelines are envisaged as streamlines of the fluid}. Crucially, the aforementioned assumption leads to $\Sigma\equiv\sigma_{ab}n^{a}n^{b}=-\Theta/3$, and therefore to $\Theta_{ab}n^{a}n^{b}=0$ ($B^{a}=\mathcal{B}n^{a}$, always). Does the condition $\Theta_{ab}n^{a}n^{b}=0$ necessarily mean that there is no collapse along the direction $n^{a}$? In accordance with the spirit of the above described analogy, our answer is negative. Instead, it is physically expected that while a given volume surrounding the magnetic forcelines does not change, its individual, $1+2$ dimensions do change. In detail, we expect that every $2$-D surface orthogonal to $n^{a}$, experiences contraction while the magnetic forcelines themselves increase in length! Finally, does this mean that the magnetic length increases unceasingly? The answer is given within the following subsections.\\
Overall, it seems that $u'_{a}=n^{b}{\rm D}_{b}u_{a}=0$ is an assumption imposed by the problem itself, namely by the elastic behaviour of the magnetised fluid along direction $n^{a}\parallel B^{a}$.

\subsection{Magnetic fracture limit and gravitational contraction}\label{ssec:fr-limit-contraction}

In the first place, we claim that the fracture limit must correspond to a maximum of the magnetic deformation with respect to proper time. However, given that as the deformation of an elastic medium increases, so do the internal tension stresses acting against it; it often happens that the maximum deformation coincides with the maximum resisting tension stress (see the following subsection~\ref{ssec:further-frac-limit}). In reference to our case, because the law of magnetic elasticity, eq.~\eqref{el-law2}, is valid at all times during the collapse (for any deformation), it is necessary\footnote{The Alfv$\acute{e}$n speed, $c^{2}_{\mathcal{A}}=1/(\beta+1)$ with $\beta\equiv (\rho+P)/\mathcal{B}^{2}$, is also a monotonically increasing function of proper time during gravitational contraction. This happens because the magnetic density, $\mathcal{B}^{2}\propto a^{-6}$ (refer to~\cite{MT}), with $a$ being the fluid's scale factor, increases faster than the matter density and pressure under a polytropic or barotropic equation of state (see the following subsection and appendix~\ref{appC}).} that the fracture limit corresponds to a double maximum, of the magnetic deformation and the magneto-curvature tension as well (i.e. $\dot{u}_{\text{fr}}=0=\dot{s}^{*}_{\text{fr}}$)\footnote{The monotonic increase of $u$ and $s^{*}$ during contraction is given by the problem's nature.}. Besides, a maximum of spatial deformation along a direction $n^{a}$, i.e. $u=-\mathcal{R}_{ab}n^{a}n^{b}$, does not make physical sense by itself within the problem of magnetised gravitational collapse. The quantity $u$ should increase monotonically without ever reaching a maximum value. It is only through the magnetic field presence (with its elastic behaviour encoded by~\eqref{el-law2}), appearing in $s^{*}$ but not directly in $u$, that a maximum of the latter acquires physical beingness together with a parallel maximum of the former. Beyond that maximum value the fieldlines of the magnetised fluid are expected to be broken, so that eventually and discontinuously $s^{*}$ becomes zero. Within the present subsection we consider the implications of the condition $\dot{u}_{\text{fr}}=0$. In the following subsection we proceed to the full determination of the magnetic fracture limit.

Thus, in mathematical terms, we proceed to the differentiation of~$u\equiv-\mathcal{R}_{ab}n^{a}n^{b}$ with respect to proper time (associated with the fluid's motion), which leads to
\begin{equation}
    \dot{u}=-\frac{2}{3}\dot{\rho}-\dot{\mathcal{E}}=\frac{1}{2}\Theta(\rho+P+\mathcal{B}^{2}+3\mathcal{E})\,,
    \label{dot-u}
\end{equation}
assuming an ideal fluid model and homogeneity of implosion. In particular, on deriving the above we have taken into account the continuity equation as well as the propagation equation for the tidal tensor, projected twice along the magnetic direction, namely (see Appendix~\ref{appC})
\begin{equation}
    \dot{\rho}=-\Theta(\rho+P)\hspace{15mm} \text{and}\hspace{15mm} \dot{\mathcal{E}}=-\frac{3}{2}\Theta\mathcal{E}+\frac{1}{6}\Theta(\rho+P-3\mathcal{B}^{2})\,.
\end{equation}
Note that within the framework of magnetised gravitational contraction we require the satisfaction of the following conditions (i.e. positive curvature magnetic deformation and increasing magnetic distortions)
\begin{equation}
    u<0\Leftrightarrow \mathcal{E}>-\frac{2}{3}\rho\hspace{15mm} \text{and} \hspace{15mm} \dot{u}>0\Leftrightarrow \mathcal{E}>-\frac{1}{3}(\rho+P+\mathcal{B}^{2})\,.
    \label{eq:contraction-conditions}
\end{equation}
Subsequently, setting eq.~\eqref{dot-u} equal to zero, we find out that the function's critical point corresponds to
\begin{equation}
    \mathcal{E}_{\text{fr}}=-\frac{1}{3}(\rho+P+\mathcal{B}^{2})_{\text{fr}}<0\hspace{15mm} \text{and}\hspace{15mm} u_{\text{fr}}=-\frac{1}{3}(\rho-P-\mathcal{B}^{2})_{\text{fr}}\,,
    \label{frac-limit}
\end{equation}
where the negative sign requirement of $u_{\text{fr}}$ implies the condition $\rho_{\text{fr}}>P_{\text{fr}}+\mathcal{B}^{2}_{\text{fr}}$. Essentially, the critical point in question is determined by the values of the matter and magnetic density (a relation between $P$ and $\rho$ can always be assumed through an equation of state, see subsection~\ref{ssec:further-frac-limit}).
Now in reference to the problem of magnetised gravitational collapse we face the following question. \textit{Do magnetic fieldlines affect the fate of gravitational collapse? In particular, will they manage to impede contraction towards a singularity or will they be inevitably broken beforehand?} Our experience shows that under increasing external tension elastic media are ultimately broken. Let us examine the case of magnetic elasticity under increasing gravity. Envisaged under a cause-consequence (causal) perspective, the magneto-curvature tension $c^{2}_{\mathcal{A}}\mathcal{R}_{ab}n^{a}n^{b}$ is the exclusive result of the system's gravity $R_{ab}u^{a}u^{b}$ (recall eq~\eqref{Raych-mod})\footnote{Besides, during the principally relativistic phenomenon of collapse, gravity is plausibly pointed out as the dominant term, cause of magnetic deformation.}. Hence, our argument is the following: If $s^{*}$ turns out to be smaller than the fluid's gravity at the fracture limit, then it should have also been smaller earlier during the collapse. In such a case magnetic fieldlines are not able to impede contraction towards a singularity (refer also to~\cite{MT}). In other words, our collapse criterion reads
\begin{equation}
        s^{*}=-(c^{2}_{\mathcal{A}}u)_{\text{fr}}<(R_{ab}u^{a}u^{b})_{\text{fr}}\,,
    \label{eq:grav-coll-cond}
\end{equation}
Although we have not yet fully determined the magnetic fracture limit, by comparing the aforementioned relativistic terms at the critical point~\eqref{frac-limit} we explicitly arrive at (recall that $0<c^{2}_{\mathcal{A}}<1$)
\begin{equation}
    \frac{c^{2}_{\mathcal{A}\text{(fr)}}}{3}(\rho-P-\mathcal{B}^{2})_{\text{fr}}<\frac{1}{2}(\rho+3P+\mathcal{B}^{2})_{\text{fr}}\,.
    \label{frac-collapse}
\end{equation}
We observe that the criterion~\eqref{eq:grav-coll-cond} is clearly satisfied at the limit and it was consequently satisfied earlier. Therefore, the magnetic forcelines are expected to reach their fracture limit before managing to impede the contraction. It is worth noting that we have reached the aforementioned conclusion without yet assuming a specific equation of state. Our result is in fact based on the assumptions of an ideal MHD fluid model and of its homogeneous contraction. In the following subsection it becomes evident that condition~\eqref{frac-collapse} is satisfied at the fully determined fracture limit.

\subsection{Complete specification of the magnetic fracture limit--Applications to neutron stars and white dwarfs}\label{ssec:further-frac-limit}

In the present subsection we fully determine the magnetic fracture limit as a double maximum, of the magnetic deformation $u$ and the magneto-curvature tension stress $s^{*}$ as well. In practice, taking the dot derivative of~\eqref{el-law2} under the condition $\dot{u}_{\text{fr}}=0$ we get
\begin{equation}
\dot{\left(s^{*}\right)}_{\text{fr}}=\dot{\left(c^{2}_{\mathcal{A}}\right)}_{\text{fr}}=-\frac{1}{\mathcal{B}^{2}(1+\beta)^{2}}\left[\dot{\rho}+\dot{P}-\frac{2}{\mathcal{B}}\dot{\mathcal{B}}(\rho+P)\right]_{\text{fr}}\,,
    \label{eq:dot-c}
\end{equation}
where $\beta\equiv(\rho+P)/\mathcal{B}^{2}$ (see appendix~\ref{appD} for details). Assuming a polytropic equation of state (i.e. $P=k\rho^{\gamma}$, where $k$ and $\gamma$ are constant parameters), and setting~\eqref{eq:dot-c} equal to zero we arrive at the condition
\begin{equation}
    \gamma P^{2}+(\gamma-1)\rho P-\rho^{2}=0\,.
    \label{eq:P-condition}
\end{equation}
Note that in deriving the above we have taken into account the propagation equation for the magnetic field at the magnetohydrodynamic limit, namely $\dot{\mathcal{B}}=-\Theta\mathcal{B}$ (see~\cite{MT}). Moreover, it is simply verified that expression~\eqref{eq:dot-c} is positive for $P<\rho/\gamma$, and negative in the opposite case. The real solutions of the above quadratic are\footnote{It is easy to check that $c^{2}_{\mathcal{A}}$ is an increasing function of proper time ($\Theta<0$ is always assumed) for $P<\rho/\gamma$ whilst decreasing for $P>\rho/\gamma$.}
\begin{equation}
    P_{\text{fr}}=k\rho^{\gamma}_{\text{fr}}=\frac{\rho_{\text{fr}}}{\gamma}\Leftrightarrow\rho_{\text{fr}}=\left(\frac{1}{\gamma k}\right)^{\frac{1}{\gamma-1}}\hspace{15mm} \text{and}\hspace{15mm} P_{\text{fr}}=-\rho_{\text{fr}}\,,
    \label{eq:quadr-sols}
\end{equation}
of which we accept the former and reject the latter, under the consideration of ordinary collapsing matter. Note that eq~(\ref{eq:P-condition}a) provides the values (in geometrised units) of matter density and pressure at the magnetic fracture limit. Overall, we have associated the magnetic fracture limit with the unique (double) maximum of $u$ and $s^{*}$ with respect to proper time. The aforementioned point (a theoretical approach to the magnetic fracture limit) is certainly expected to slightly differ, in practice, from the actual instant in which the magnetic fieldlines are broken (i.e. $s^{*}=0$).\\
We can make use of the information provided by~(\ref{eq:quadr-sols}a), together with an initial setting, in order to predict how much has the volume of a given magnetised (collapsing) star changed until reaching its magnetic fracture limit. In particular, with the aid of~\eqref{density-evol-polytrope} (see appendix~\ref{appC}), we deduce that the relation between matter density/scale factor at the fracture limit (i.e. $\rho_{\text{fr}}$ and $a_{\text{fr}}$) and their counterparts at an initial (hydrostatic equilibrium) state of the star (i.e. $\rho_{0}$ and $a_{0}$), is
\begin{equation}
    \rho_{\text{fr}}=\rho_{0}\left(\frac{\frac{C}{k}a^{2}_{0}-1}{\frac{C}{k}a^{2}_{\text{fr}}-1}\right)^{\frac{1}{\gamma-1}}\hspace{15mm} \text{or} \hspace{15mm} a^{2}_{\text{fr}}=\frac{k}{C}\left[\left(\frac{\rho_{0}}{\rho_{\text{fr}}}\right)^{\gamma-1}\left(\frac{C}{k}a^{2}_{0}-1\right)+1\right]\,.
    \label{eq:fr-0}
\end{equation}
The above formula points out the sensible conclusion that the more dense a star is, the earlier it reaches its fracture limit. Furthermore, substituting~(\ref{eq:quadr-sols}a) into~\eqref{frac-collapse} we deduce once again (formally this time) that the magnetic forcelines are broken before managing to impede contraction. In the following we consider, as an application, two illustrative examples of a neutron star and a white dwarf.

\subsubsection{Neutron star of mass $M=1.5M_{\odot}$, radius $R=10$~km and average magnetic field $\mathcal{B}\sim 10^{12}$~G}\label{sssec:neutron-star}

A neutron star with the above mentioned characteristics has matter and magnetic densities $\rho\sim 10^{-14}~\text{cm}^{-2}$ and $\mathcal{B}^{2}\sim 10^{-27}~\text{cm}^{-2}$, in geometrised units (i.e. $c=1=G$. For conversion between cgs and geometrised units see e.g. the appendix of~\cite{JBH}). Assuming that the matter of the star consists in average of non-relativistic neutrons, the parameters of its polytropic equation of state are $\gamma=5/3$ and $k=5.3802\cdot 10^{9}~$cgs$~\sim 10^{6.7}~\text{cm}^{4/3}$ (e.g. refer to~\cite{ST} or~\cite{JSB} for details regarding the parameters' values). Finally, from~\eqref{density-evol-polytrope} written in the initial conditions we determine the value of the integration constant, $C=(\rho^{2/3}_{0}+k)/a^{2}_{0}\sim 10^{9.3}a^{-2}_{0}$. Taking into account that $\mathcal{B}^{2}\propto a^{-6}$ (refer to~\cite{MT}), and substituting all the aforementioned values in~(\ref{eq:fr-0}b) we find out that
\begin{equation}
    a_{\text{fr}}\sim 10^{-0.7}a_{0}\sim 10^{-1}a_{0}\,,\hspace{3mm} \rho_{\text{fr}}\sim 10^{2}\rho_{0}\sim10^{-12}~\text{cm}^{-2} \hspace{3mm}\text{and}\hspace{3mm} \mathcal{B}^{2}_{\text{fr}}\sim 10^{6}~\mathcal{B}^{2}_{0}\sim 10^{-21}~\text{cm}^{-2}\,.
    \label{eq:neutron-star-fract}
\end{equation}
Therefore, we expect that the magnetic forcelines will be broken when the neutron star's radius becomes about ten times smaller (that is $1$~km) than its initial (equilibrium) value. In parallel, the magnetic/matter density ratio will have grown by four orders of magnitude, i.e. $(\mathcal{B}^{2}/\rho)_{\text{fr}}\sim 10^{4}~(\mathcal{B}^{2}/\rho)_{0}\sim 10^{-9}$.

\subsubsection{White dwarf of mass $M=0.6M_{\odot}$, radius $R=1.4\cdot 10^{-2}$~$R_{\odot}$ and average magnetic field $\mathcal{B}\sim 10^{6}$~G}\label{sssec:white-dwarf}

The white dwarf in question has matter and magnetic densities $\rho\sim 10^{-23}~\text{cm}^{-2}$ and $\mathcal{B}^{2}\sim 10^{-39}~\text{cm}^{-2}$, in geometrised units. Assuming that the stellar fluid mainly consists of ultra relativistic electrons, the parameters of its polytropic equation of state read $\gamma=4/3$ and $k\sim 10^{15}$~cgs$\sim 10^{4.3}~\text{cm}^{2/3}$. Under the above mentioned initial conditions the integration constant of~\eqref{density-evol-polytrope} gives $C\sim 10^{7.6}~a^{-2}_{0}$. Overall, (recalling once again that $\mathcal{B}^{2}\propto a^{-6}$) we find out that
\begin{equation}
    a_{\text{fr}}\sim 10^{-4.8}a_{0}\sim 10^{-5}a_{0}\,,\hspace{3mm} \rho_{\text{fr}}\sim 10^{28}\rho_{0}\sim 10^{5}~\text{cm}^{-2} \hspace{3mm}\text{and}\hspace{3mm} \mathcal{B}^{2}_{\text{fr}}\sim 10^{28.8}~\mathcal{B}^{2}_{0}\sim 10^{-10}~\text{cm}^{-2}\,.
    \label{eq:white-dwarf-fract}
\end{equation}
We deduce that the stellar fluid reaches its magnetic fracture limit when its radius shrinks to approximately one hundred thousand times its initial value (that is the fracture radius is some hundreds of meters). As for the magnetic/matter density ratio, it increases by about an order of magnitude, i.e. $(\mathcal{B}^{2}/\rho)_{\text{fr}}\sim 10^{0.8}~(\mathcal{B}^{2}/\rho)_{0}\sim 10^{-15}$.

\section{Discussion}

Overall, the essence of our reasoning lies in that gravitational deformation of magnetic forcelines is governed by Hooke's law of elasticity, originating from the Ricci identities. However, there are two basic features distinguishing gravitational distortions of magnetic forcelines from mechanical distortions of elastic materials. Firstly, unlike mechanical distortions of elastic materials, Hooke's law in the form of~\eqref{el-law2} is not an approximate expression only valid for small magnetic deformations (thus magnetic forcelines do not seem to have an elastic limit). In contrast, as long as Ricci identities are an appropriate definition of spatial curvature for large values of the latter (advanced stages of gravitational collapse), the law in question consists of an exact expression, valid for any size of distortion. Secondly, the proportionality factor in the elasticity law~\eqref{el-law2} is a variable instead of a constant quantity.

Based on the aforementioned law we have calculated the magnetic fracture limit under gravitational volume distortions, and subsequently applied it in two explicit cases of a neutron star and a white dwarf. Considering our results as new, we raise the fundamental problem regarding the observational-experimental (and further theoretical) verification of the gravito-magnetic elasticity law. Although magnetic elasticity under great gravitational distortions is practically not a subject offered for study in earthly laboratories, progress towards the experimental path could alternatively and in the first place be achieved by examining magnetic distortions under progressively increasing rotations. Finally, focussing on the magnetic reaction against gravity, the present work peaks with the argument that magnetic forcelines reach their fracture limit before managing to impede gravitational contraction. Interestingly, knowing how the magnetic fieldlines explicitly behave during gravitational implosion, could hopefully shed light on the various evolution phases of astrophysical objects. The motivation for the above sentence essentially comes from considering that many stars or protogalactic clouds are associated with (even small) magnetic fields which are rapidly increasing during their collapse.

Wherever they appear in our universe, either in astrophysical or in cosmological environments, magnetic fields are impelled by gravity to manifest their extraordinary elastic features. The phenomena arising from those properties await our exploration.\\

\appendix

\section{Propagation equations and constraints for the kinematically induced magnetic tension stresses}\label{appA}

In order to arrive at the propagation equation for $\sigma^{(B)}_{ab}$, we make the following steps. First, project eq.~\eqref{4-Ricci} along the timelike 4-velocity $u^{a}$; second, project orthogonal to $u^{a}$ with the aid of $h_{ab}$ and with respect to both indices (removing thus timelike terms); third, take the symmetric and trace-free part of the resulting relation. The equation in question finally reads\footnote{An index with bar denotes that the associated component has been projected orthogonal to $u^{a}$.}
\begin{eqnarray}
    \dot{\sigma}^{(B)}_{\bar{a}\bar{b}}&=&-\Theta\dot{u}_{\langle a}B_{b\rangle}+2\dot{u}_{\langle a}\omega_{b\rangle}{}^{c}B_{c}+{\rm D}_{\langle a}\dot{B}_{\bar{b}\rangle}-\left(\sigma_{c\langle a}+\omega_{c\langle a}+\frac{1}{3}\Theta h_{c\langle a}\right)\sigma^{c(B)}{}_{b\rangle}\nonumber\\
    &&-\left(\sigma_{c\langle a}+\omega_{c\langle a}\right)\omega^{c(B)}{}_{b\rangle}-\frac{1}{2}B_{\langle a}q_{b\rangle}\,,
    \label{prop-B-shear}
\end{eqnarray}
where $\dot{u}_{a}=u^{b}\nabla_{b}u_{a}$ is the fluid's acceleration and $q_{a}$ its flux vector. On deriving the above we have taken into account eq 1.3.1 of~\cite{TCM}, as well as that
\begin{equation}
    \nabla_{a}B_{b}={\rm D}_{a}B_{b}-u_{a}\dot{B}_{b}+u_{b}(\nabla_{a}u^{d})B_{d}+u_{a}u_{b}\dot{u}^{c}B_{c} \hspace{5mm} \text{and} \hspace{5mm} R_{a\langle bc\rangle d}u^{a}B^{d}=\frac{1}{2}B_{\langle b}q_{c\rangle}\,,
\end{equation}
where eqs 1.2.6, 1.2.8 and 1.2.11 of~\cite{TCM} have been used on finding the latter of the above. Following a similar procedure but taking the antisymmetric part of~\eqref{4-Ricci} (via contraction with the 3-D Levi-Civita pseudotensor $\epsilon_{abc}$) this time, we arrive at the propagation equation for the magnetic tension induced by twisting effects
\begin{eqnarray}
    \dot{\omega}^{(B)}_{\bar{a}}&=&-3\epsilon_{abc}\dot{u}^{b}\sigma^{c}{}_{d}B^{d}-\epsilon_{abc}{\rm D}^{b}\dot{B}^{\bar{c}}-\epsilon_{abc}\left(\sigma^{bd}+\omega^{db}+\frac{1}{3}\Theta h^{bd}\right){\rm D}_{d}B^{c}\nonumber\\
    &&-(\dot{u}^{b}B_{b})\omega_{a}+(\dot{u}^{b}\omega_{b})B_{a}+H_{ab}B^{b}-\frac{1}{2}\epsilon_{abc}B^{b}q^{c}\,,
    \label{prop-B-vorticity}
\end{eqnarray}
where $H_{ab}$ is the magnetic Weyl component and we have taken into account that
\begin{equation}
    \dot{B}_{\bar{a}}=-\frac{2}{3}\Theta B_{a}+(\sigma_{ab}+\epsilon_{abc}\omega^{c})B^{b}\hspace{5mm} \text{and} \hspace{5mm} \epsilon_{abc}R^{ebcd}u_{e}B_{d}=H_{ab}B^{b}-\frac{1}{2}\epsilon_{abc}B^{b}q^{c}\,.
    \label{Farad}
\end{equation}
Note that eq~(\ref{Farad}a) is an expression of Faraday's law at the MHD limit. On the other hand, the spacelike part of~\eqref{4-Ricci} leads to the divergence conditions for the aforementioned quantities. In detail, we start from the 3-D Ricci identities~\eqref{3-Ricci}. Subsequently, we take either its trace or its contraction with $\epsilon_{abc}$. The former case leads to
\begin{equation}
    {\rm D}^{b}\sigma^{(B)}_{ab}=\text{curl}\omega^{(B)}_{a}+2\omega_{ab}\left(-\frac{2}{3}\Theta B^{b}+\sigma^{b}{}_{c}B^{c}\right)+\mu\omega_{a}-2\omega^{2}B_{a}+\mathcal{R}_{ba}B^{b}
    \label{Div-B-shear}
\end{equation}
whilst the latter to
\begin{equation}
    {\rm D}^{a}\omega^{(B)}_{a}=\frac{1}{6}\Theta\mu-2\sigma_{ab}\omega^{a}B^{b}\,.
    \label{Div-B-vorticity}
\end{equation}
In~\eqref{Div-B-shear} $\mathcal{R}_{ab}$ represents the 3-D Ricci tensor.
On deriving the above we have made use of~(\ref{Farad}a) as well as of 
\begin{equation}
    \omega^{a}B_{a}=\mu/2 \hspace{5mm} \text{and} \hspace{5mm} \epsilon^{abc}\mathcal{R}_{dcba}B^{d}=-\frac{2}{3}\Theta\mu-4\sigma_{ab}\omega^{a}B^{b}\,,
    \label{Gauss}
\end{equation}
where $\mu$ is the charge density and~(\ref{Gauss}a) is an expression of Gauss's law at the MHD limit. It is worth noting that for zero rotational distortions (i.e. $\omega_{ab}=0$) of the magnetic field eqs~\eqref{prop-B-shear} and~\eqref{Div-B-shear} significantly simplify to\footnote{An ideal fluid (i.e. $q_{a}=0$) has been assumed in the first equation.}
\begin{equation}
    \dot{\sigma}^{(B)}_{\bar{a}\bar{b}}=-\Theta\dot{u}_{\langle a}B_{b\rangle}+{\rm D}_{\langle a}\dot{B}_{\bar{b}\rangle}-\left(\sigma_{c\langle a}+\frac{1}{3}\Theta h_{c\langle a}\right)\sigma^{c(B)}{}_{b\rangle}\hspace{5mm} \text{and} \hspace{5mm} {\rm D}^{b}\sigma^{(B)}_{ab}=\mathcal{R}_{ba}B^{b}\,.
\end{equation}
Overall, equations~\eqref{prop-B-shear},~\eqref{prop-B-vorticity},~\eqref{Div-B-shear} and~\eqref{Div-B-vorticity} determine the kinematics of the magnetic tension stresses triggered by shear and vorticity effects.

\section{Our approach to magnetised contracting flow in technical terms}\label{appB-after}

\footnote{The present appendix section is envisaged as a correction of Appendix A in~\cite{MT}. In particular, we point out here that eqs (102) and (106) of the aforementioned work are valid, not generally, but within a specific framework.}The present appendix unit provides a technical supplement to subsection~\ref{ssec:Approach-magn-grav-collapse}, within the main text. Throughout the present manuscript we encounter several times products of tensors with the spacelike (unit) vector field $n^{a}$ (such that $n_{a}u^{a}=0$ and $n_{a}n^{a}=1$), which is taken parallel to the magnetic forcelines (i.e. $B^{a}=\mathcal{B}n^{a}$). In particular, our calculations often involve the quantities $\Sigma=\sigma_{ab}n^{a}n^{b}$ and $\Sigma_{a}=\tilde{h}_{a}{}^{b}\sigma_{bc}n^{c}$ (with $\tilde{h}_{ab}=g_{ab}+u_{a}u_{b}-n_{a}n_{b}=h_{ab}-n_{a}n_{b}$, satisfying $\tilde{h}_{ab}n^{b}=0$). Concerning the former, it can be written as
\begin{equation}
\Sigma\equiv\sigma_{ab}n^{a}n^{b}\equiv {\rm D}_{\langle b}u_{a\rangle}n^{a}n^{b}={\rm D}_{(b}u_{a)}n^{a}n^{b}-\frac{1}{3}\Theta h_{ab}n^{a}n^{b}=u'_{a}n^{a}-\frac{1}{3}\Theta\,,
\label{eqn:appB1}
\end{equation}
where $u'_{a}=n^{b}{\rm D}_{b}u_{a}$. Note that the term $u'_{a}n^{a}$ is not generally zero because the prime ($'$) denotes an actually spatial derivative operator. This means that $'$ does not generally satisfy the product (Leibniz) rule between (the timelike) $u^{a}$ and (the spacelike) $n^{a}$. Our approach consists thus of setting $u'_{a}n^{a}$ equal to zero. The quantity in question vanishes either when prime differentiation obeys the product rule\footnote{Let us consider firstly the decomposition
${\rm D}_{b}n_{a}=h_{b}{}^{c}h_{a}{}^{d}\nabla_{c}n_{d}=\nabla_{b}n_{a}+u_{b}\dot{n}_{a}-u_{a}({\rm D}_{b}u_{c})n^{c}$.
Projecting the above along $n^{b}$ and recalling that ${\rm D}_{b}u_{a}=\sigma_{ab}+\omega_{ab}+(\Theta/3)h_{ab}$, we receive
\begin{equation}
n^{b}{\rm D}_{b}n_{a}=n^{b}\nabla_{b}n_{a}-u_{a}(n^{b}\nabla_{b}u_{c})n^{c}=n^{b}\nabla_{b}n_{a}-\left(\Sigma+\frac{\Theta}{3}\right)u_{a}\,.\nonumber
\end{equation}
We observe that under the condition $\Sigma=-\frac{\Theta}{3}$, the above
equation recasts into $n^{b}\nabla_{b}n_{a}=n^{b}{\rm D}_{b}n_{a}$~(e1). Secondly, recalling $\nabla_{b}u_{a}={\rm D}_{b}u_{a}-u_{a}\dot{u}_{b}$, we observe that $n^{b}\nabla_{b}u_{a}=n^{b}{\rm D}_{b}u_{a}$~(e2) always. Overall, it is clear that via expressions~(e1), (e2), the prime derivative obeys the product rule between $u^{a}$ and $n^{a}$. In other words, we deduce that (see also our initial eq.~\eqref{eqn:appB1})
\begin{equation}
u'_{a}n^{a}=(u_{a}n^{a})'-u_{a}n'^{a}=0\hspace{5mm} \text{if and only if}\hspace{5mm} \Sigma=-\frac{\Theta}{3}\,.\nonumber
\end{equation}
Note that the last conclusion is not directly implied by~\eqref{eqn:appB1}. In fact, the latter, under condition~\eqref{eqn:app-Σ}, leads to $u'_{a}n^{a}=0$ but not necessarily to Leibniz's rule for prime differentiation.} or when $u'_{a}=n^{b}{\rm D}_{b}u_{a}=0$. Hence, we have
\begin{equation}
\Sigma=-\frac{\Theta}{3}\hspace{8mm}\text{for}\hspace{8mm} u'_{a}=0\,.
\label{eqn:app-Σ}
\end{equation}
\textit{The last condition implies that the fluid velocity is homogeneous along the magnetic forcelines. In other words, our approximation is equivalent with considering the magnetic forcelines as streamlines of the fluid}.\\
Besides, it is worth noting that the double projection of the volume expansion/contraction tensor, $\Theta_{ab}=\sigma_{ab}+(\Theta/3)h_{ab}$, along a spatial direction $n^{a}$ gives
\begin{equation}
\Theta_{ab}n^{a}n^{b}=u'_{a}n^{a}\,.
\label{eqn:theta-tensor}
\end{equation}
Interestingly, taking into account~\eqref{eqn:app-Σ}, the above expression becomes $\Theta_{ab}n^{a}n^{b}=0$. Crucially, in our problem of magnetised gravitational collapse (i.e. $B^{a}=\mathcal{B}n^{a}$), $\Theta_{ab}n^{a}n^{b}=0$ translates into the following physical assumption. During contraction, a given volume containing a bundle of magnetic fieldlines, does not change. As an illustrative instance, let us consider a cylindrical such volume. Any (circular) cross section of the latter decreases during implosion, whilst its length (i.e. the length of the magnetic forcelines) increases (so that the cylindrical volume remains constant). \textit{Does this mean that the magnetic length increases incessantly to infinity? Our answer to the problem is definitely no}. Instead, we claim that there is a magnetic fracture limit (in the sense of elastic dynamics), which we calculate within the main text.\\\\
In reference to the other shear component of interest, namely $\Sigma_{a}$, we have
\begin{equation}
\Sigma_{a}\equiv \tilde{h}_{a}{}^{b}\sigma_{bc}n^{c}\equiv \tilde{h}_{a}{}^{b}n^{c}{\rm D}_{\langle c}u_{b\rangle}=\tilde{h}_{a}{}^{b}n^{c}{\rm D}_{(c}u_{b)}=\frac{1}{2}\Sigma_{a}+\frac{1}{2}\epsilon_{ab}\Omega^{b}+\frac{1}{2}\tilde{h}_{a}{}^{b}u'_{b}\,,
\end{equation}
which subsequently leads to
\begin{equation}
\Sigma_{a}=\epsilon_{ab}\Omega^{b}+\tilde{h}_{a}{}^{b}u'_{b}\hspace{8mm}\text{or}\hspace{8mm} \Sigma_{a}=\epsilon_{ab}\Omega^{b}\hspace{5mm}\text{for}\hspace{5mm} u'_{a}=0\,.
\label{eqn:Sigma-vector}
\end{equation}
Moreover, we take into account that the projection of Faraday's law (within ideal MHD-see eq. (41) in~\cite{MT}) normal to the magnetic direction $n^{a}$, gives
\begin{equation}
\Sigma_{a}=\epsilon_{ab}\Omega^{b}+\alpha_{a}\,,
\end{equation}
where $\alpha_{a}\equiv\tilde{h}_{a}{}^{b}\dot{n}_{b}$. Comparing the above with~(\ref{eqn:Sigma-vector}a), we find out that
\begin{equation}
\alpha_{a}=\tilde{h}_{a}{}^{b}u'_{b}\hspace{8mm}\text{and}\hspace{8mm}\alpha_{a}=0\hspace{5mm}\text{for}\hspace{5mm} u'_{a}=0\,.
\label{eqn:alpha-u'}
\end{equation}
The last expression will prove useful in appendix~\ref{appC}.
We make use of eqs~\eqref{eqn:app-Σ} and~\eqref{eqn:Sigma-vector} in various calculations throughout this manuscript. 

\section{Magneto-curvature tension stresses}\label{appB}

Following discussion in subsection~\ref{ssec:magneto-curvature tension} the magneto-curvature tension stresses associated with shear, rotational and volume curvature distortions are
\footnote{On deriving eqs~\eqref{s-STF}-\eqref{elast-law1} we make use of the so-called Gauss-Codacci formula (e.g. see eq. 1.3.39 in~\cite{TCM}).}
\begin{eqnarray}
    s_{\langle ac\rangle}&=&\mathcal{B}^{2}\mathcal{R}_{d\langle ac\rangle b}n^{b}n^{d}=\nonumber\\
    &&\mathcal{B}^{2}\left[\epsilon_{\langle a|q|}\epsilon_{c\rangle s}\mathcal{E}^{qs}+\frac{1}{2}\left(2\Pi n_{\langle a}n_{c\rangle}+2\Pi_{\langle a}n_{c\rangle}-\pi_{ac}\right)-\frac{\Theta}{3}\left(\Sigma n_{\langle a}n_{c\rangle}+4\Sigma_{\langle a}n_{c\rangle}\right)\right]\,,
    \label{s-STF}
\end{eqnarray}
\begin{equation}
    s_{[ac]}=\mathcal{B}^{2}\mathcal{R}_{d[ac]b}n^{b}n^{d}=\mathcal{B}^{2}\left(\Pi_{[a}n_{c]}-\frac{4\Theta}{3}\Sigma_{[a}n_{c]}\right)
    \label{twisting}
\end{equation}
and
\begin{equation}
    s=s^{c}{}_{c}=\mathcal{B}^{2}\mathcal{R}_{bd}n^{b}n^{d}=\mathcal{B}^{2}\left(\frac{2}{3}\rho+\mathcal{E}+\Pi\right)\,,
    \label{appeq:elast-law1}
\end{equation}
where $\pi_{ab}$ and $E_{ab}$ are the anisotropic stress and the tidal (or electric Weyl) tensors respectively. Moreover, we have $\Pi\equiv \pi_{ab}n^{a}n^{b}$, $\Pi_{a}\equiv \tilde{h}_{a}{}^{b}n^{c}\pi_{bc}$, $\mathcal{E}\equiv E_{ab}n^{a}n^{b}$, $\mathcal{E}_{ab}\equiv (\tilde{h}_{(a}{}^{c}\tilde{h}_{b)}{}^{d}-(1/2)\tilde{h}_{ab}\tilde{h}^{cd})E_{cd}$, $\Sigma\equiv \sigma_{ab}n^{a}n^{b}=-\Theta/3$ and $\Sigma_{a}\equiv \tilde{h}_{a}{}^{b}n^{c}\sigma_{bc}=\epsilon_{ab}\Omega^{b}$ (see appendix~\ref{appB-after} for the last two expressions), with $\epsilon_{ab}\equiv \epsilon_{abc}n^{c}$ being the 2-D counterpart of the Levi-Civita pseudotensor, $\Omega_{a}\equiv\tilde{h}_{a}{}^{b}\omega_{b}$ and $\tilde{h}_{ab}\equiv h_{ab}-n_{a}n_{b}$ an operator projecting orthogonal to the magnetic field direction $n^{a}$. We observe that tidal effects (electric Weyl components) are associated with shape and volume magnetic distortions only. Assuming an ideal fluid model, the anisotropic stress terms in the above vanish. Then, of particular interest is that the deformation due to gravitational compression/expansion in~\eqref{appeq:elast-law1} is determined by the density  of matter and the tidal tensor projected along the magnetic fieldlines.

\section{a) Temporal evolution of the matter density under a polytropic equation of state\, b) Propagation equation for tidal stresses along the magnetic forcelines}\label{appC}

Considering an ideal, polytropic (i.e. $P=k\rho^{\gamma}$, with $k$ and $\gamma$ constants) fluid at the MHD limit, the continuity equation, $\dot{\rho}=-\Theta(\rho+P)$ ($\Theta=3\dot{a}/a$, $a$ denoting the scale-factor of the fluid's volume), reads the following explicit Bernoulli form
\begin{equation}
    \frac{d\rho}{da}+\frac{3}{a}\rho+\frac{3k}{a}\rho^{\gamma}=0\,.
    \label{cont-eq-polytrope}
\end{equation}
The equation in question accepts the general solution
\begin{equation}
    \rho=\left[Ca^{-3(1-\gamma)}-k\right]^{\frac{1}{1-\gamma}}\,,
    \label{density-evol-polytrope}
\end{equation}
with $C$ (note that $C>0$ for $k>0$) being the integration constant. In the cases of non-relativistic neutrons ($\gamma=5/3$) and ultra-relativistic electrons ($\gamma=4/3$) the above equation recasts into
\begin{equation}
    \rho=\left(Ca^{2}-k\right)^{-3/2}\hspace{15mm} \text{and}\hspace{15mm} \rho=\left(Ca-k\right)^{-3}
    \label{density-evol-polytrope2}
\end{equation}
respectively. Obviously, the pressure of matter, $P=k\rho^{\gamma}$, increases faster than its density for $\gamma>1$ (i.e. the cases we consider). We employ the above equations in determining the magnetic fracture limit of a neutron star and a white dwarf in the main text (see subsection~\ref{ssec:further-frac-limit}).\\
In the following second part of the present appendix unit we present some details regarding the propagation equation for $\mathcal{E}$ used in subsection~\ref{ssec:fr-limit-contraction}.

In particular, for homogeneous collapse of a magnetised ideal fluid the propagation equation for the tidal tensor reads
\begin{eqnarray}
    \dot{E}_{\langle ab\rangle}&=&-\Theta E_{ab}-\frac{1}{2}(\rho+P)\sigma_{ab}-\frac{1}{2}\dot{\pi}_{ab}-\frac{1}{6}\Theta\pi_{ab}+3\sigma_{\langle a}{}^{c}\left(E_{b\rangle c}-\frac{1}{6}\pi_{b\rangle c}\right) \nonumber\\
    &&+\epsilon_{cd\langle a}\left[2\dot{u}^{c}H_{b\rangle}{}^{d}-\omega^{c}\left(E_{b\rangle}{}^{d}+\frac{1}{2}\pi_{b\rangle}{}^{d} \right) \right]\,,
    \label{dot-Eab}
\end{eqnarray}
where $H_{ab}$ is the so-called magnetic Weyl component and $\pi_{ab}$ is exclusively sourced from the magnetic field. Projecting the above twice along the magnetic direction $n^{a}$ we ultimately arrive at
\begin{equation}
    \dot{\mathcal{E}}=-\frac{3}{2}\Theta\mathcal{E}+\frac{1}{6}\Theta(\rho+P)-\frac{1}{2}\Theta B^{2}\,.
    \label{dot-E}
\end{equation}
It is worth noting that in deriving the above we have taken into account that $\Sigma=-\Theta/3$ and $\Sigma_{a}=\epsilon_{ab}\Omega^{b}$ (where for the inner product of two arbitrary vectors $k^{a}$ and $l^{a}$ with the $2-D$ Levi-Civita tensor we have $\epsilon_{ab}k^{a}l^{b}=0$). Moreover, we have made use of $E_{ab}\dot{n}^{a}n^{b}=\mathcal{E}_{a}\alpha^{a}=0$, where $\alpha_{a}\equiv \tilde{h}_{a}{}^{b}\dot{n}_{b}$ (see eq.~\eqref{eqn:alpha-u'}).

\section{Some auxiliary calculations}\label{appD}

In reference to eq~\eqref{eq:dot-c} (i.e. temporal derivative of the Alfv$\acute{\text{e}}$n speed) in subsection~\ref{ssec:further-frac-limit}, we employ the continuity equation and the equation of state ($P=k\rho^{\gamma}$) to calculate the dot derivative of matter density and pressure, i.e. $\dot{\rho}=-\Theta(\rho+P)$ and $\dot{P}=\gamma(P/\rho)\dot{\rho}$. Furthermore, we make use of the law of magnetic contraction, $\dot{\mathcal{B}}=-\Theta\mathcal{B}$ (under the ideal MHD approximation of a magnetised fluid, and condition~\eqref{eqn:app-Σ}), introduced and described in~\cite{MT}. Therefore, eq~\eqref{eq:dot-c} recasts into
\begin{equation}
    \dot{\left(c^{2}_{\mathcal{A}}\right)}_{\text{fr}}=-\frac{1}{\mathcal{B}^{2}(1+\beta)^{2}}\left[\dot{\rho}+\dot{P}-\frac{2}{\mathcal{B}}\dot{\mathcal{B}}(\rho+P)\right]=-\frac{\Theta(1+\beta)^{2}}{\mathcal{B}^{2}}\left[\rho+(1-\gamma)P-\gamma\frac{P^{2}}{\rho}\right]=0\,,
    \label{eq:c-dot-app1}
\end{equation}
which clearly leads to~\eqref{eq:P-condition}. Alternatively, assuming a barotropic equation of state (i.e. $P=w\rho$ with $w=\text{constant}$), eq~\eqref{eq:dot-c} transforms into
\begin{equation}
    \dot{\left(c^{2}_{\mathcal{A}}\right)}_{\text{fr}}=-\frac{1}{\mathcal{B}^{2}(1+\beta)^{2}}\left[\dot{\rho}+\dot{P}-\frac{2}{\mathcal{B}}\dot{\mathcal{B}}(\rho+P)\right]=-\frac{\Theta\rho}{\mathcal{B}^{2}(1+\beta)^{2}}(1-w^{2})=0\,,
    \label{eq:c-dot-app2}
\end{equation}
from which we deduce that $w=\pm 1$ or $P=\pm\rho$. The negative solution is directly rejected (ordinary collapsing matter is assumed) whilst the negative one leads to $u_{\text{fr}}=-(1/3)(\rho-P-\mathcal{B}^{2})_{\text{fr}}=\mathcal{B}^{2}/3>0$ (see eq~\eqref{frac-limit}), also not accepted because $u$ must be negative at all times during the collapse. Hence, it seems that a barotropic equation of state is not appropriate for describing magnetised gravitational collapse, of neutrons stars or white dwarfs.

Finally, we mention here for reference the values of the second temporal derivative of $u$ and the first temporal derivative of $\mathcal{E}$ at the fracture limit. In detail, the dot differentiation of~\eqref{dot-E} under the condition~\eqref{frac-limit}, as well as $\dot{\rho}=-\Theta(\rho+P)$ and $\dot{\mathcal{B}}=-\Theta\mathcal{B}$, leads to
\begin{equation}
\dot{\mathcal{E}}_{\text{fr}}=-\frac{1}{2}\Theta\left(3\dot{\mathcal{E}}-\frac{\dot{\rho}+\dot{P}}{3}+2\mathcal{B}\dot{\mathcal{B}}\right)_{\text{fr}}=\frac{2}{3}\Theta(\rho+P)_{\text{fr}}<0\,,
\label{eq:dot-E-fr}
\end{equation}
where $\Theta<0$ (for contraction), and $\dot{\mathcal{E}}_{\text{fr}}=-(2/3)\dot{\rho}_{\text{fr}}$ is verified by definition of $u$ (i.e. $\equiv-(2/3)\rho-\mathcal{E}$). Similarly, the proper-time differentiation of~\eqref{dot-u} under the condition~\eqref{frac-limit}, as well as the aforementioned propagation formulae, $P=k\rho^{\gamma}$ and eq~\eqref{eq:quadr-sols}, leads to
\begin{equation}
\ddot{u}_{\text{fr}}=\frac{1}{2}\Theta\left(3\dot{\mathcal{E}}+\dot{\rho}+\dot{P}+2\mathcal{B}\dot{\mathcal{B}}\right)_{\text{fr}}=-\left(\Theta^{2}\mathcal{B}^{2}\right)_{\text{fr}}<0\,.
\label{ddot-u1}
\end{equation}
\\

\textbf{Acknowledgements:} The present research work was supported by the Hellenic Foundation for Research and Innovation (H.F.R.I.), under the ‘Third Call for H.F.R.I. PhD Fellowships’ (Fellowship No. 74191 ). Earlier partial support from the Foundation for Education and European Culture is also acknowledged. I would finally like to thank Professor Christos G. Tsagas for his advise and comments; as well as Professors Charalambos Moustakidis and Georgios Pappas for sharing with me their knowledge, useful information and references.\\

\end{document}